\documentclass[prb,twocolumn,amsmath,amssymb]{revtex4}
\usepackage{graphicx}
\begin{document}

%\preprint{}

\title{Tunnelling-charging Hamiltonian of a Cooper pair pump
at large $E_{\mathrm{J}}/E_{\mathrm{C}}$:\\ Modified Hamiltonians
and renormalisability}


\author{M. Aunola}
\email[]{Matias.Aunola@phys.jyu.fi}
%\thanks{}
\altaffiliation{Finnish Defence Forces Technical Research Centre (PvTTei-os), PL 10,
FIN-11311 Riihim\"aki, FINLAND}
\affiliation{Department of Physics, University of Jyv\"askyl\"a,
P.O. Box 35 (YFL), FIN-40014 University of Jyv\"askyl\"a, FINLAND}


\date{\today}

\begin{abstract}
The properties of the tunnelling-charging Hamiltonian of a Cooper pair
pump are well understood in the regime of weak and intermediate
Josephson coupling, i.e.  $E_{\mathrm{J}}\lesssim E_{\mathrm{C}}$.
Instead of perturbative treatment of charging effects, the present
work applies the charge state representation in the the strong
coupling case. From the discrete Hamiltonian we construct effective,
truncated PDE Hamiltonians and analytically obtain approximate
ground-state wave functions and eigenenergies. The validity of the
expressions is confirmed by direct comparison against the results of
numerical diagonalisation. For uniform arrays, our results converge
rapidly and even $\phi$-dependence of the wave function is described
reasonably.  In the inhomogeneous case we find the Hamiltonian to be
parametrically renormalisable. A method for finding inhomogeneous
trial wave function is explained. The intertwined connection linking
the pumped charge and the Berry's phase is explained, too. As addendum,
we have explicitly validated the ground state ansatz for $\phi=0$
when $N\le 42$.
\end{abstract}

%\pacs{}
%\keywords{}

\maketitle

\section{Introduction}
Josephson junction devices, e.g. Cooper pair boxes,
superconducting single electron  transistors (SSET)
and  Cooper pair pumps,  have been extensively studied
both theoretically\cite{ave98,pek99,mak99,ave00,fal00,pek01} and 
experimentally.\cite{moo99,nak99,orl99,cho01,bib02} 
For a recent review, see Ref.~\onlinecite{mak01}.
Possible applications include at least
direct Cooper pair pumping,\cite{pek99} decoherence 
studies,\cite{pek01} related metrological applications,\cite{has99} 
and the use of Cooper pair charge qubits or persistent-current qubits 
(SQubits) in quantum computation.\cite{ave98,moo99}

The ideal tunnelling-charging Hamiltonian of a Cooper pair pump has 
been studied in detail in Refs.~\onlinecite{pek99,aun00,aun01}.
Charge transfer due to direct supercurrent and adiabatic
pumping due to varying gate voltages have been adequately 
described when the Josephson coupling is weak or at most
comparable to charging effects. The case of strong 
Josehpson coupling in ideally biased arrays is still
relatively unexplored. A single Josephson junction
is known to be described by the Mathieu equation\cite{abr72}
in the phase representation. For a superconducting single electron 
transistor (SSET) the charge state representation is identical
to one-dimensional discrete harmonic oscillator and, thus, 
the Mathieu equation in the island's phase 
representation.\cite{eil94,tink96} 

In this paper we first develop a method for obtaining
an approximate solution of the Mathieu equation. Later on, 
we generalise the method for several dimensions and
make the required corrections for our model Hamiltonian.
In short, starting from the discrete Hamiltonian we construct a
modified partial differential equation (PDE) for which
a trial solution is obtained. Subsequently, the solution is
overlaid as the wave function the discrete Hamiltonian and
the result is compared against numerically obtained 
eigenstate. 

In order to sum up the obtained results we state the following:
For homogeneous arrays of arbitrary length 
we find analytical and rapidly converging
wave functions and eigenenergies. These expressions are
derived from the developed method  
The case of non-zero phase difference is treated in a fairly
satisfactory way. Inhomogeneous arrays are first treated
by parametric renormalisation which yields an accurate 
approximation for the ground state energy. A modification
of the original method improves the wave function, but 
not the asymptotical rate of convergence.

Skeel and Hardy\cite{ske01} have performed
analysis on constructing modified Hamiltonian when integrating
systems of PDE's over time, see also Refs.~\onlinecite{rei99,gan00,hai00}.
In these works numerical discretisation is approximately counteracted
by using a suitable truncation of the modified equations. 
The principles of the present method are similar,
although it is applied on a discrete eigenvalue problem.

This paper is organised as follows. In Sec.~\ref{sec:hamil}
the Hamiltonian is defined and its structure is explained.
In Sec.~\ref{sec:method} we find an approximate solution
for the Mathieu equation in charge state representation
and postulate the generalisation of the method for several
coordinates. In Sec.~\ref{sec:strong} homogeneous arrays are examined
and explicit trial wave functions for the ground state
are constructed.
In Sec.~\ref{sec:phidep} the developed formalism is 
extended to into account non-zero values of phase difference
across the array. In Sec.~\ref{sec:inhomog} the Hamiltonian
is shown to be parametrically renormalisable in the
inhomogeneous case. Wave function is also constructed 
although the accuracy is not as good as in
the homogeneous case. Finally, the conclusions
are drawn in Sec.~\ref{sec:conclu}.

\section{Constructing the Hamiltonian
\label{sec:hamil}}

\begin{figure}[htb]
\includegraphics[width=55truemm]{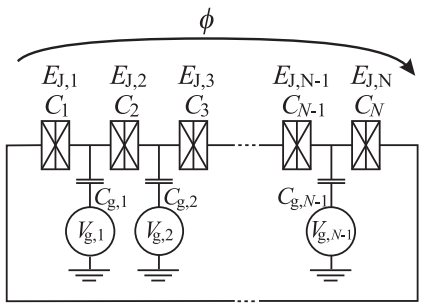}
\caption{{}An ideal superconducting array of independent
Josephson junctions. Here $C_k$ and $E_{\mathrm{J},k}$ are the
capacitance and the Josephson energy of the $k^{\mathrm{th}}$ junction,
respectively. The total phase difference across the array, $\phi$,
is a constant of motion.
\label{fig:pump}}
\end{figure}

A schematic view of the system is shown in
Fig.~\ref{fig:pump}. We assume that the gate voltages
$V_{\mathrm{g},j}$ are independent and externally operated.
The bias voltage across the array, $V_{\mathrm{b}}$,
which controls the total phase difference $\phi$
according to  $d\phi/dt=-2eV/\hbar$, is assumed to
be ideally set to zero. Hence, $\phi$ remains fixed and 
becomes a good quantum number in proper variables,
which have been presented e.g.~in Ref.~\onlinecite{ing92}. 
On the other hand, a precise value of $\phi$ means that 
its conjugate variable $\hat M$, the average number of
tunnelled Cooper pairs ($\hat M:=-i\partial/\partial\phi$), 
becomes completely undetermined.

In the following, the 
tunnelling-charging Hamiltonian
\begin{equation}
H=H_{\mathrm{C}}+H_{\mathrm{J}},\label{eq:simphami}
\end{equation}
is assumed to be the correct description of the microscopic system.
The ideal model Hamiltonian simply neglects quasiparticle tunnelling
as well as other degrees of freedom. The most important 
parameters are the (average) Josephson coupling energy
$E_{\mathrm{J}}$ and (average) charging energy, defined
as $E_{\mathrm{C}}:=(2e)^2/(2C)$. These determine ``the 
Josephson-charging ratio'' which is denoted by
$\varepsilon_{\rm J}:=E_{\mathrm{J}}/E_{\mathrm{C}}$.
The Hamiltonian and the operation of a Cooper pair pump 
in the weak coupling regime is further 
determined by (normalised) gate charges 
$\vec q:=\{q_1,\ldots,q_{N-1}\}$, where
$q_k:=-C_{{\rm g},k} V_{{\rm g},k}/2e$. In the present model
relative junction capacitances $\vec c$, where
$c_k:=C_k/C$ and $\sum_{k=1}^NC_k^{-1}=N/C$, also determine
individual Josephson energies by $E_{\mathrm{J},k}:=c_kE_{\mathrm{J}}$.
For uniform or homogeneous arrays we have 
$c_k:=1$, while the inhomogeneity can be reliably
quantified by the inhomogeneity index $X_{\rm inh}:=
[\sum_k (c_k^{-1}-1)^2/N]^{1/2}$.\cite{aun00,aun01}

The matrix elements of the charging Hamiltonian $H_{\rm C}$ are given by 
the capacitive charging energy and thus they read\cite{aun00}
\begin{equation}
\langle \vec n\vert H_{\rm C}(\vec q)\vert \vec n\rangle_{\phi}=
E_{\rm C}\left[\sum_{k=1}^N\frac{v_k^2}{c_k}
-\frac1N\left(\sum_{k=1}^N\frac{v_k}{c_k}\right)^2\right],
\label{generchar0}
\end{equation}
where the number of Cooper pairs on each island is given 
by $\vec n=(n_1,\ldots,n_{N-1})$. 
The quantities $\{v_k\}_{k=1}^N$  are an arbitrary solution of 
the charge conserving equations
\begin{equation}
v_k-v_{k+1}=n_k-q_k.\label{eq:veeesi}
\end{equation}
Tunnelling of one Cooper pair through the $k$th
junction changes $\vert\vec n\rangle$ by $\vec\delta_k$,
where the non-zero components are (if
applicable) $(\vec \delta_k)_k=1$ and $(\vec \delta_k)_{k-1}=-1$.
The tunnelling Hamiltonian is given by
\begin{equation}
H_{\rm J}=-\sum_{\vec n,k=1}^{N}\frac{c_kE_{{\rm J}}}{2}(\vert
\vec n+\vec\delta_k\rangle\langle \vec n\vert e^{i\phi/N}+\,{\rm H.c.}\,).
\label{eq:tunnel}
\end{equation}
The supercurrent flowing through the array is determined by
the supercurrent operator $I_{\rm S}=(-2e/\hbar)(\partial H
/\partial \phi)$,
a G$\hat{\mathrm a}$teaux derivative\cite{rud73} of the
full Hamiltonian. By changing the gate voltages adiabatically along a
closed path $\Gamma$, a charge transfer
$Q_{\rm tot}:=Q_{\rm s}+Q_{\rm p}$ is induced.
The pumped charge, $Q_{\rm p}$, depends only on
the chosen path, while the charge due to direct supercurrent, $Q_{\rm s}$,
also depends on how the  gate voltages are operated.
If the system remains in a
adiabatically evolving state $\vert m\rangle$, the total transferred 
charge, $Q_{\rm tot}$, in units of $-2e$, reads\cite{pek99,aun01}
\begin{eqnarray}
%\frac1{\hbar}\int_0^\tau
%\frac{\partial E_m(t)}{\partial \phi}dt+
-\frac{\partial \eta_m(t)}{\partial \phi}+
2\oint_{\Gamma}{\rm Re}\left[\langle m\vert\hat M
\vert dm\rangle\right].
\label{eq:simppump}
\end{eqnarray}
where $\vert dm\rangle$ is the
change in $\vert m\rangle$ due to a differential change of the gate
voltages $d\vec q$ and $\eta_m=-\int_0^\tau (E_m(t)/\hbar)dt$
is the dynamical phase of the wave function.

Clearly, the pumped charge is closely related to the 
the geometrical Berry's phase,\cite{ber84}
$\gamma_m(\Gamma)=i\oint_{\Gamma} \langle m\vert dm\rangle$.
The pumped charge can be evaluated from Eq.~(\ref{eq:simppump})
in the charge state representation 
once the overall phase of the eigenstate is fixed consistently
for all $\vec q$.
If the examined state is sufficiently non-degenerate for 
all values of $\phi$, the eigenstate can be expanded as a 
Fourier series in $\phi$ with real coefficients $\{a_{\vec n,l}\}$.
Consequently, for a fixed value of $\phi$ the differential pumped charge 
is given by a gauge-invariant expression\cite{aun01}
\begin{eqnarray}
dQ_{\rm p}(\phi)\hskip-2pt\!&=&\!\hskip-3pt
\sum_{l'=0}^\infty\sum_{\vec n,l=-\infty}^\infty
\hskip-1pt\left[
\frac{2(l+Y_{\vec n}/N)}{1+\delta_{l'0}}d(a_{\vec n,l}a_{\vec n,l+l'})
\right.\cr
&&\ \!\hskip-5pt+\left. l'(a_{\vec n,l}da_{\vec n,l+l'}\hskip-2pt-
\hskip-2pt a_{\vec n,l+l'}da_{\vec n,l})\right]\cos(l'\phi),
\label{eq:transchar}
\end{eqnarray}
where $Y_{\vec n}$ is an additional class label.
In constrast, a differential change in the phase difference $\phi$
for fixed gate charges $\vec q$ 
induces no pumped charge, because we find
\begin{equation}
dQ_p(d\phi)=2{\rm Im}\left[\langle \hat M m\vert
\hat M m\rangle\right] d\phi=0.
\end{equation}

\begin{figure}[htb]
\includegraphics[width=70truemm]{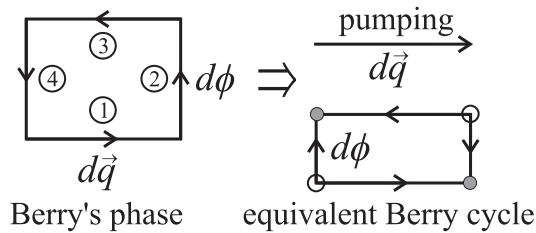}
\caption{{}An infinitesimal cycle $C$ corresponding to Berry's
phase $\gamma_m(C)$ consists of four legs. The charge transfer
$Q_{\rm p}$ for a fixed $\phi$ is identical to the Berry's
phase induced by traversing the legs in the shown directions.
This path can not be followed continuously in the
$(\vec q,\phi)$-plane.  \label{fig:cycle}}
\end{figure}

Now consider the Berry's phase $\gamma_{m}$ induced by an 
infinitesimal closed cycle $C$ at $(\vec q,\phi)$ with 
sides $d\vec q$ and $d\phi$ as shown by the l.h.s.~pf 
Fig.~\ref{fig:cycle}.
The result divided by $d\phi$, i.e. $d\gamma_m^{(C)}/d\phi$,
is identical to $dQ_{\rm p}$ apart from the sign of the first term.
In other words, the contribution from the first and third part
of the cycle gives the non-integrable part of $dQ_{\rm p}$,
while the second and fourth part add up to the integrable part
multiplied by $-1$. Thus the path for which the $\phi$-''derivative''
of Berry's phase is identical to $dQ_{\rm p}(\phi)$ is not a closed
cycle but a more complex path illustrated in the r.h.s.~of 
Fig.~\ref{fig:cycle}.

From here on the expression for the charging energy, 
Eq.~(\ref{generchar0}),
is examined in detail. This is done in order to rewrite the
Hamiltonian in as simple a form as possible.
In the homogeneous case the quadratic
form is easily diagonalised and we find $N-1$ identical
eigenvalues of $E_{\mathrm{C}}$ and one zero-energy mode
in the direction of $\hat v_0:=(1,1,\ldots,1)/\sqrt{N}$.
This demonstrates the uniqueness of the charging
energy expression for each charge state and, consequently,
the same zero-energy mode is observed in the inhomogeneous 
case, too.

%If the array is uniform, the charging energy reads
%$$
%E_{\mathrm{ch}}^{\vec v}=0(\tilde v_0)^2+
%E_{\mathrm{C}}\sum_{j=1}^{N-1}(\tilde v_j)^2,
%$$
%where $\{\tilde v_j\}_{j=0}^{N-1}:=\{\hat v_j\cdot \vec v
%\}_{j=0}^{N-1}$ is the component along the axis $\hat v_j$
%and $\{\hat v_j\}$ define an orthonormal coordinate system
%of the $\hat v$-space. The vectors
%\begin{equation}
%\hat v_{j}
%:=(1,e^{i(2\pi j/N)},e^{i(4\pi j/N)},\ldots,
%e^{-i(2\pi j/N)})/\sqrt{N},\label{eq:veevekt}
%\end{equation}
%where $j=0,\ldots,N-1$ or, more appropriately, normalised, 
%linearly independent real and imaginary parts of $\hat v_j$
%provide such a basis.

In a proper representation of
the $q$-space, the charging energy for homogeneous arrays
can be expressed as $E_{\mathrm{C}}
\Vert\tilde q\Vert_2^2$, where $\Vert\, \cdot\,\Vert_2$
is the usual Euclidean norm.  Thus,
the representatives of the tunnelling vectors 
$\{\vec \delta_j\}$, denoted by $\hat q_j$, are required.
Above all, they must be normalised according to
\begin{equation}
\hat q_j\cdot\hat q_k=\delta_{jk}-1/N.\label{eq:normalise}
\end{equation}
In an orthonormal $(N-1)$-dimensional 
basis, where $\hat e_j\cdot\hat e_k=\delta_{jk}$ and
$\vec x=(x_1,x_2,\ldots,x_{N-1})$,
the representatives define variables $\{\tilde q_j\}$ 
according to
\begin{equation}
\tilde q_j(\vec x):=\sum_{k=1}^{N-1}( e_k\cdot q_j)x_k.
\label{eq:coordinates}
\end{equation}
The normalisation condition~(\ref{eq:normalise}) yields
relations
\begin{equation}
\sum_{j=1}^N\tilde q_j=0\quad\mathrm{and}
\quad \sum_{j=1}^N\tilde q_j^2=\sum_{j=1}^{N-1}x_j^2=\Vert \vec x\Vert^2
\label{eq:qrel}
\end{equation}
which are valid for all values of $\vec x$.
%In principle, one could now obtain the representatives
%from Eq.~(\ref{eq:veevekt}), because an orthonormal 
%basis transformation is always unitary, and component related to 
%the zero-energy mode is simply removed. Unfortunately, the
%obtained representatives are not very convenient.

\begin{figure}[htb]
\vspace{0.2truecm}
\includegraphics[width=70truemm]{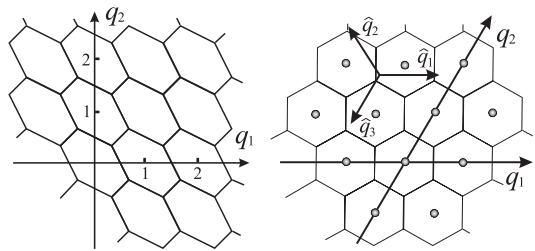}
\caption{{} On the left-hand-side,
the so-called honey comb structure 
induced by the charge state lattice ($N=3$).
The regular lattice determined by the representatives $\hat q_1$, 
$\hat q_2$, and $\hat q_3$ is shown on the right-hand side.
The origin of a new rectangular coordinate system $(x_1,x_2)$
is set by the gate charges $\vec q=(q_1,q_2)$.
The charging energy for a charge state (gray circle) then reads
$E_{\mathrm{C}}(x_1^2+x_2^2)$,
where the nearest-neighbour distance has been scaled to $\sqrt{2/3}$.
\label{fig:honey}}
\end{figure}

Suitable representatives for cases $N=3$ and $N=4$ are easy to find
and their visualisation is obvious. When $N=3$, we select
\begin{eqnarray}
\hat q_1&=&(\sqrt{2/3},0),\ \ \hat q_2=(-1/\sqrt{6},1/\sqrt{2}),\cr
\hat q_3&=&(-1/\sqrt{6},-1/\sqrt{2}),\label{eq:repN3}
\end{eqnarray}
which describes three directions separated by identical $120^{\circ}$
angles. 
The resulting transformation of coordinates and the so-called 
honeycomb structure is shown in Fig.~\ref{fig:honey}. 
The gate charges $q_1$ and $q_2$ determine the origin of
the induced, rectangular coordinate system $(x_1,x_2)$.

For $N=4$,  symmetric representatives are given
by the well-known body centered cubic lattice (BCC) 
of solid state physics, explicitly
\begin{eqnarray}
\hat q_1&=&(1,1,-1)/2,\ \ \hat q_2=(1,-1,1)/2,\cr
\hat q_3&=&(-1,1,1)/2,\ \ \hat q_4=(-1,-1,-1)/2.\label{eq:BCCrep}
\end{eqnarray}
These representatives are convenient when studying the case $N=4$, but
a more general method for obtaining representaves is required. By
augmenting the existing representatives for $N$ we can always obtain
the set for $N+1$. The additional representative is set to lie along
the new (first) coordinate axis with the correct length
$\sqrt{N/(N+1)}$.  The normalisation condition~(\ref{eq:normalise}) is
satisfied if all other representatives are retained as they were with
an identical first component of $-1/\sqrt{N(N+1)}$.

Applying this method inductively, starting from trivial case of $N=1$,
yields the general representives for any $N$. Let the length of the
array be $N$ and denote the $j^{\mathrm{th}}$ representive and its
$k^{\mathrm{th}}$ component by $\hat q^N_j$ and $q^N_{j(k)}$,
respectively.  The $N-1$ components are obtained from three 
simple rules:
\begin{eqnarray}
&(i)& q^N_{j(j)}=\sqrt{(N-j)/(N+1-j)}\cr
&(ii)& q^N_{j(k>j)}=0\cr
&(iii)& q^N_{j(k<j)}=-1/\sqrt{(N-j)(N+1-j)}\label{eq:qrules}
\end{eqnarray} 
The above transformation simplifies and symmetrices 
the tunnelling-charging  Hamiltonian for arrays of any 
length. Inhomogeneous arrays
can also be considered once the tools have been developed.

\section{Mathieu equation and discrete harmonic oscillator\label{sec:method}}

The canonical form of the Mathieu equation reads\cite{abr72}
\begin{equation}
\frac{d^2y}{dv^2}+(a-2q\cos(2v))y=0,
\end{equation}
where $y(v)$ is the solution, $q$ is a parameter and 
$a$ is known as the characteristic 
value or eigenvalue. 

The Hamiltonian of a SSET for a fixed phase difference
$\phi$ in the charge state representation 
can be mapped onto a one dimensional discrete harmonic 
oscillator (DHO), see e.g. Ref.~\onlinecite{tink96,aun02}.
Our chosen form includes a nearest-neighbour 
coupling $-\varepsilon_{\mathrm{J}}/2$ and 
the potential $V(n)=(n-n_0)^2$, where $n$ is an integer.
The equation for the amplitude  $a_n$ now reads
\begin{equation}
(n-n_0)^2a_n-(\varepsilon_{\mathrm{J}}/2)(a_{n-1}+a_{n+1})=E a_n,
\label{eq:discho}
\end{equation}
where  $E$ is the eigenvalue we are looking for.
In order to obtain the solution of the discrete equation,
we assume that $a_n$ is a continuous function an replace
other amplitudes by respective Taylor expansions. We denote the
step size by $h$ (here $h=1$) which yields
\begin{equation}
a_{n-h}+a_{n+h}=\sum_{k=0}^{\infty}\frac{2h^{2k}}{(2k)!}\frac{d^{2k}
a_n}{d n^{2k}},\label{eq:derivatives}
\end{equation}
a differential equation for $a_n$. 

We now transform into conjugate variables of the island charge,
i.e. $n-n_0\rightarrow -id/d\theta$ and 
$id/dn\rightarrow \theta$. Collecting the terms, we find
\begin{equation}
\frac{d^2 a(\theta)}{d\theta^2}+\left[E+
(\varepsilon_{\mathrm{J}}/2)\cos(\theta)\right]a(\theta)=0,
\end{equation}
which is identical to the Mathieu equation with $a=4E$ and 
$q=2\varepsilon_{\mathrm{J}}$ once we choose
$\tilde\theta=(\theta+\pi)/2$. 

In the limit $q\rightarrow\infty$, the ground state energy 
can be read from Eq.~20.2.30 of Ref.~\onlinecite{abr72}
with the result
\begin{equation}
E_0^{\mathrm{DHO}}\hskip-2pt=\hskip-1pt-\varepsilon_{\mathrm{J}}
\hskip-1pt+\hskip-1pt\frac{\sqrt{\varepsilon_{\mathrm{J}}}}{\sqrt2}
\hskip-1pt-\hskip-1pt\frac{1}{16}\hskip-1pt-\hskip-1pt
\frac{\sqrt{2/\varepsilon_{
\mathrm{J}}}}{256}\hskip-1pt-\hskip-1pt\frac{3/\varepsilon_{\mathrm{J}}}{2048}
\hskip-1pt+\hskip-1pt
\mathcal{O}(\varepsilon_{\mathrm{J}}^{-3/2}),\label{eq:dhoene}
\end{equation}
confirmed by numerical diagonalisation, too.
Returning to the charge state representation, we divide 
the eigenvalue problem by $\varepsilon_{\mathrm{J}}$ and define
the oscillator frequency $\omega:=\sqrt{2/\varepsilon_{\mathrm{J}}}$
and scaled energy $\tilde E:=E/\varepsilon_{\mathrm{J}}+1$.
The lowest  order approximation becomes
\begin{equation}
-a''/2+\mbox{$\frac12$}\omega^2(n-n_0)^2a=\tilde E,
\end{equation}
which is analytically solvable with $\tilde E=\omega/2$
and $a(n)\propto \exp(-\omega(n-n_0)^2/2)$.
From here on, $n_0$ is omitted from the expression $(n-n_0)$
for brevity.

The discretisation naturally affects the wave function
and as well as the eigenenergy~(\ref{eq:dhoene}).
The lowest order approximation $\psi_1$ for the discrete wave
function $\psi_d$ is naturally a Gaussian wave function.
The optimal, but unnormalised, wave function is given by
\begin{equation}
\psi_{1}(n)\propto \exp\left(-\frac{\omega n^2/2}{1-\omega/8}\right).
\label{eq:simpwfn}
\end{equation}
%
% Next the first excited state
%
\begin{equation}
\psi_{1}^{\mathrm{ex}}(n)
\propto n\exp\left(-\frac{\omega n^2/2}{1-27\omega/16}\right).
\end{equation}
More accurate a wave function reads
\begin{equation}
\psi_{2}(n)\propto \exp\left(-\frac{\omega n^2(1-\omega^2n^2/
48)/2}{1-3\omega/16}\right),\label{eq:advwfn}
\end{equation}
%
% and the excited state (1/ej convergence at least up to  ~10^4)
%
\begin{equation}
\psi_{2}^{\mathrm{ex}}(n)\propto \left(n-\frac{\omega^2n^3}{24}
\right)\exp\left(-\frac{\omega (n^2-\omega^2n^4/
48)/2}{1-147\omega/640}\right),
\end{equation}
where a cutoff $(1-\omega^2n^2/48)\rightarrow 1$ must be
applied for large enough values of $n$, when the deviation
becomes greater than $20$--$30$ \%.    
These wave functions have been compared against the 
result of numerical diagonalisation, $\psi_d$, by taking
the norm of the difference,
in short $\Vert \psi_d-\psi_j\Vert$, which yields approximately
\begin{eqnarray}
\Vert \psi_d-\psi_1\Vert&\approx& 0.018/\sqrt{\varepsilon_{\mathrm{J}}},\\
\Vert \psi_d-\psi_2\Vert&\approx& 0.009/\varepsilon_{\mathrm{J}}.
\end{eqnarray}
Because both trial wave functions converge towards the actual eigenstate
of the system, approximate eigenenergies corresponding to $\psi_1$ and
$\psi_2$ can be easily evaluated. Setting $n_0=0$ and examining the 
equation for coefficient $a_0$ gives  
\begin{equation}
E_{\psi_j}\sim  -\varepsilon_{\mathrm{J}}(a_1/a_0),
\label{eq:energyvalue}
\end{equation}
where $\varepsilon_{\mathrm{J}}\rightarrow\infty$.
Expanding the terms in powers of $\omega=\sqrt{2/\varepsilon_{\mathrm{J}}}$
gives the desired result. We find that $E_{\psi_1}$ first deviates from
the constant order in which the term is $-1/8$ instead of the correct
$-1/16$. As expected $E_{\psi_2}$ is much better, and even the term
$-\sqrt{2/\varepsilon_{\mathrm{J}}}/256$ is correctly reproduced.

The significance of the corrections in $\psi_2(n)$ with respect to 
the continuous solutions is relatively clear. 
The denominator $1-3\omega/16$  cancels $3/4$ of the of the 
leading second order term $-(\omega/2)^2/2$ and gives the 
correct eigenenergy in the constant order. On the other hand,
the term proportional to $n^4$ is related to the truncated differential
operator
\begin{equation}
\frac{d^2}{dn^2}+\frac1{12}\frac{d^4}{dn^4}.
\end{equation}
The coefficient $-\omega^2/48$ can be divided in two parts, namely
$1/12$ and $-\omega^2/4$ which seems reasonable as the latter scales
correctly as function of $\omega$, while the former changes  if the
step length $h$ is altered.  

This approach is rather similar to that of Skeel and
Hardy\cite{ske01} although they consider time-dependent
problems instead of eigenvalue problems. Systems of differential
equations are replaced by modified equations which try to
compensate for the discretisation error. 
The present potential is harmonic and
in the conjugate representation truncated potentials are
anharmonic in nature.  Anharmonic oscillators have been studied,
and exact eigenvalues have been obtained.\cite{mei99,meu02}
Unfortunately, the sign of our leading correction is negative,
so these works are not applicable here.

If there are two orthogonal and independent directions, 
the wave function factorises and the one-dimensional result can be 
generalised. Nevertheless, we make the following assumption which is
to be justified later. Let our Hamiltonian be defined on a 
regular, discrete lattice of the coordinates $\vec x$ and the
potential be isotropic and harmonic, i.e.
$V(\vec x)=\omega^2\Vert \vec x\Vert^2/2$. Interactions between
(neigbouring) lattice sites are expanded in terms of partial
derivates up the fourth order  in a 
similar manner to Eq.~(\ref{eq:derivatives}).
We postulate an analytical trial solution if 
the second order operator is the Laplacian and
the modified PDE eigenvalue problem has the form
\begin{equation}
-\mbox{$\frac12$}\left(\nabla^2+D_4/12\right)\psi+\mbox{$\frac12$}\omega^2
\Vert \vec x\Vert^2\psi=\tilde E\psi,\label{eq:modified}
\end{equation}
where $D_4$ is a fourth order partial differential operator.
We define corresponding 'conjugate variable' $\tilde D_4$ by 
replacing each partial derivative with respect to $x_j$ by 
$x_j$ itself. For example, if $D_4=(\nabla^2)^2$, i.e.
the square of the Laplacian, the conjugate variable is the
fourth power of the norm, explicitly,
$\tilde D_4=(\Vert \vec x\Vert^2)^2=\Vert \vec x\Vert^4$.
Our unnormalised trial wave function is given by
\begin{equation}
\psi_2(\vec x)\propto \exp\left(\frac{-(\omega/2)(\Vert \vec x\Vert^2
-\omega^2\tilde D_4/48)}{1-\alpha \omega}\right),
\end{equation} 
where $\alpha$ is chosen so that cancellation of the constant
order term in energy is exactly $3/4$, just as the factor $3/16$ in
Eq.~(\ref{eq:advwfn}). The conjugate variable $\tilde D_4$ gives
the correct functional form, although a cutoff for too large
values as compared to $\Vert \vec x\Vert^2$ must be naturally applied.
Hopefully, the asymptotic 
convergence of the norm $\Vert \psi_d-\psi_2\Vert$ is 
better than $1/\sqrt{\varepsilon_{\mathrm{J}}}$. The general
asymptotical solution for the discretised harmonic oscillator
has geen recently given in Ref.~\onlinecite{aun03}.

\section{Strong Josephson coupling and homogeneous arrays at $\phi=0$
\label{sec:strong}}

When the Josephson energy $E_{\mathrm{J}}$ is large 
as compared to charging energy $E_{\mathrm{C}}$ it seems
preferable to express the Hamiltonian in terms of the 
phase differences $\phi_j$. We choose to remain in
the charge state representation for two reasons. First,
the charging Hamiltonian is difficult to evaluate in
the independent phase representation. Additionally, the model
Hamiltonian is already diagonal with respect to the
total phase difference $\phi$. The model Hamiltonian can
be approximately diagonalised and interactions between
states with different values of $\phi$ should be included
later. 

The Hamiltonian equation is explicitly written
in units of $E_{\mathrm{C}}$, and more specifically, each 
equation (row) of the eigenvalue problem is examined separately.
Each charge state $\vec n$ is labeled according its position
in the orthonormal coordinates $\vec x=(x_1,\ldots,x_{N-1})$.
In units of $E_\mathrm{C}$,
the equation for the coefficient $a_{\vec x}$ reads
\begin{equation}
\Vert \vec x\Vert^2 a_{\vec x}-\frac{\varepsilon_{\mathrm{J}}}{2}
\sum_{j=1}^N(e^{-i\phi/N}a_{\vec x+\hat q_j}+e^{i\phi/N}
a_{\vec x-\hat q_j})=Ea_{\vec x}.
\end{equation}
Now, consider the case $\phi=0$ and large values of 
$\varepsilon_{\mathrm{J}}$ in detail. Writing the eigenvalue as
$\tilde E:=N+E/\varepsilon_{\mathrm{J}}$ transforms the 
eigenvalue problem into
\begin{equation}
-\frac12\sum_{j=1}^N(a_{\vec x-\hat q_j}\hskip-2pt-2a_{\vec x}
\hskip-2pt +\hskip-2pt a_{\vec x+
\hat q_j})\hskip-2pt+\hskip-2pt\mbox{$\frac12$}\omega^2
\Vert \vec x\Vert^2a_{\vec x}=\tilde E a_{\vec x},\label{eq:derivative0}
\end{equation}
Using the procedure explained in the previous section, we can find the
corresponding modified PDE. The truncation means that each term 
$(a_{\vec x-\hat q_j}-2a_{\vec x} +a_{\vec x+\hat q_j})$ 
corresponds to a second order derivative and a fourth order
derivative. The sum of the second order derivatives yields
the Laplacian operator $\nabla^2$ due to the second  
part of Eq.~(\ref{eq:qrel}), and the form of the modified equation 
matches Eq.~(\ref{eq:modified}).
Next, we must evaluate the form of $\tilde D_4$, find 
correct value of $\alpha$, and compare the resulting wave
function and eigenenergie against numerically obtained 
results.

The simpler, optimal Gaussian wave function reads
\begin{equation}
\psi_{1}(\vec x)\propto \exp\left(-\frac{\omega \Vert\vec x\Vert^2/2}{1-
(N-1)\omega/(8N)}\right),
\end{equation}
where the modification of $(N-1)/N$ in the denominator arises from the
fact that $\Vert\hat q_j\Vert^2=(N-1)/N$. The optimality as well as
the expected rate of convergence, i.e.
$1/\sqrt{\varepsilon_{\mathrm{J}}}$, has been confirmed up to $N=10$. 

In  case $N=3$ we find that
\begin{equation}
D_4^{(N=3)}=(\nabla^2)^2/2
\end{equation}
and hence $\tilde D_4^{(N=3)}=\Vert\vec x\Vert^4/2$. Because 
$\Vert\hat q_j\Vert^2=2/3$, the improved wave function for $N=3$
reads
\begin{equation}
\psi_2^{(N=3)}\propto \exp\left(-\frac{(\omega \Vert x\Vert^2/2)(1-\omega^2
\Vert x\Vert^2/96)}{1-\omega/8}\right).
\end{equation}
This proves to be quite accurate as the norm of the error 
vanishes according to
$\Vert \psi_d-\psi_2^{(N=3)}\Vert\approx 0.0045/\varepsilon_{\mathrm{J}}$.
As arrays become longer, pure radial (energy)
 dependence is not enough, since
the operators $D_4$ become more complicated. For the BCC representatives
($N=4$) the differential operator is given  by  
\begin{equation}
D_4:=\frac3{4}\left(\sum_{j=1}^3\frac{\partial^2}{\partial x_j^2}\right)^2
-\frac12\sum_{j=1}^3\frac{\partial^4}{\partial x_j^4},
\end{equation}
corresponding to a wave function $\psi_2^{(4)}$ proprotional to
\begin{equation}
\exp\hskip-1pt\left(
\hskip-1pt-\frac{(\omega/2)\left[\Vert x\Vert^2
\hskip-1pt-\hskip-1pt\omega^2
\left(\frac{\Vert x\Vert^4}{64}\hskip-1pt-
\hskip-1pt\sum_{j=1}^3\frac{x_j^4}{96}
\right)\right]}{1-9\omega/64}\right)\hskip-1pt.
\end{equation}
Because $\langle x_j^4\rangle=\langle \Vert\vec x\Vert^4\rangle/5$,
this also explains why the best energy dependent fit 
occurs at $3\omega\Vert\vec x\Vert^4/320$.

For longer arrays the expression for the fourth order differential
operator becomes quite complicated and less informative. Fortunately,
the value of the conjugate variable  $\tilde D_4^{(N)}$ can be easily 
obtained for any point $\vec x$. The simple expression is based on
inner product of the $\vec x$-space and the representatives
$\hat q_j$, in short
\begin{equation}
\tilde D_4^{(N)}(\vec x)=\sum_{j=1}^N(\hat q_j\cdot\vec x)^4.
\label{eq:conjugate}
\end{equation}
The differential operator $D_N$ can be read from the above expression
by retaining the components of $\vec x$ in symbolic form and
transforming each coordinate its corresponding partial derivative.
The correct cancellation requirement implies that the
general form of $\alpha$ is given by $\alpha=3(N-1)/16N$.

Thus, the general trial wave function $\psi_2^{(N)}$ becomes
\begin{equation}
\psi_2^{(N)}(\vec x)=A\exp\left(-\frac{(\omega/2)(\Vert x\Vert^2-\omega^2
\tilde D_4^{(N)}(\vec x)/48)}{1-3(N-1)\omega/(16N)}
\right),\label{eq:genwfn}
\end{equation}
where $A$ is a normalisation factor and
$\tilde D_N(\vec x)$ given in Eq.~(\ref{eq:conjugate})
is evaluated for all charge states in the used basis.
A suitable cutoff with respect to 
$\omega^2\tilde D_N(\vec x)/(48\Vert x\Vert^2)$,
e.g. between $0.2$ and $0.3$, is naturally important.  
The wave function is
independent of the  representives $\{\hat q_j\}$, but those given in 
Eq.~(\ref{eq:qrules}) are probably the most convenient.
The rate of convergence of the norm $\Vert \psi_d-\psi_2^{(N)}\Vert$
has been confirmed as $1/\varepsilon_{\mathrm J}$ up to $N=7$.
Tentatively, the same applies for  $N=10$, although
diagonalisation was limited below $\varepsilon_{\mathrm{J}}\approx 20$.

The ground state energy is 
virtually independent of the gate charges $\vec q$ when
$\varepsilon_{\mathrm{J}}$ is large enough. Thus $E_0^{(N)}$
can be  approximately obtained as  in the
one-dimensional case, see Eq.~(\ref{eq:energyvalue}).
All $2N$ neighbouring amplitudes are identical which
now gives
\begin{equation}
E_0^{(N)}\approx -N\varepsilon_{\mathrm{J}}\exp\left
[-\frac{(\omega/2)\left(\frac{N-1}{N}-\frac{\omega^2
(N-1)^2}{48N^2}\right)}{1-3(N-1)\omega/(16N)}\right].
\end{equation}
Expansion in powers of $\omega$ yields the asymptotic 
expansion
\begin{equation}
E_0^{(N)}\hskip-1pt\sim\hskip-1pt -N\varepsilon_{\mathrm{J}}
\hskip-0.5pt+\hskip-0.5pt(N-1)\sqrt{\frac
{\varepsilon_{\mathrm{J}}}{2}}\hskip-0.5pt-\hskip-0.5pt
\frac{(N-1)^2}{16N}\hskip-1pt+\hskip-0pt\mathcal{O}\hskip-2pt\left(
\varepsilon_{\mathrm{J}}^{-1/2}\right)\hskip-1pt,
\label{eq:genenergy}
\end{equation}
verified by direct comparison against the numerically obtained
eigenvalue for cases which allow diagonalisation. No analytical
expression for the term proportional to
$1/\sqrt{\varepsilon_{\mathrm{J}}}$ have been found, but it is not
correctly reproduced, either.  Direct calculation, using a method
proposed in Refs.~\onlinecite{aun03} and~\onlinecite{aun03b},
validates the above ansatz and corresponding asymptotical eigenenergy
for $N\le42$, though.\cite{teksti} We now proceed to the the case when
$\phi$ is no longer zero.

\section{Effects due to non-zero phase difference
\label{sec:phidep}}

For non-zero values of the phase difference $\phi$ the
wave function becomes complex valued because the nearest
neighbour coupling contains a term $e^{\pm i\phi/N}$. When
$\phi$ is sufficiently small the phase does not vary
significantly between nearest neighbours and as the first 
approximation the phase can be neglected in the corresponding
equations. We then consider the absolute value of the amplitudes
and observe that the differential operator is simply multiplied 
by a factor $\cos(\phi/N)$. 

Consequently, the approximate eigenvalue problem to the original one,
except that $\omega$ is replaced by $\tilde\omega=\omega/
\sqrt{\cos(\phi/N)}$. The ground state energy can be obtained from
Eq.~(\ref{eq:genenergy}) with $\varepsilon_{\mathrm{J}} \rightarrow
\varepsilon_{\mathrm{J}}\cos(\phi/N)$. The accuracy of this expression
is rather good, even for large values of $\phi$ if $
\varepsilon_{\mathrm{J}}$ is sufficiently large. The convergence in
terms of the absolute values of the amplitudes is satisfactory, too. 
Convergence in terms of trial wave function $\vert \tilde \psi_1\vert$ 
goes clearly as $1/\sqrt{\varepsilon_{\mathrm{J}}}$ 
and that of $\vert\tilde\psi_2\vert$ goes nearly
as $1/\varepsilon_{\mathrm{J}}$, weakening as
$\phi$ increases.

In order to consider the complex wave function explicitly, the
approximate differential operator induced by $\phi$ must be 
constructed. The first order differential operator is 
always cancelled on behalf of the first property in Eq.~(\ref{eq:qrel}).
The common prefactor of the third order terms,
relative to the Laplace operator, is here $-i\sin(\phi/N)/3$.
Because the conjugate coordinate $\tilde D_4^{(N)}$ was so
successful in describing the homogeneous case, we define 
a third order conjugate coordinate which evaluates to
\begin{equation}
\tilde D_3^{(N)}(\vec x)=\sum_{j=1}^N(\hat q_j\cdot\vec x)^3.
\label{eq:conjugate2}
\end{equation}
The first guess for the phase of the trial wave function is
then given by
\begin{equation}
\frac{\psi_2(\vec x)}{\vert \psi_2(\vec x)\vert}
\approx \exp\left(-\frac{(\tilde\omega/2)
(-i\tilde\omega \tilde D_N^{(3)})/(6N)}{1-3(N-1)\tilde\omega/(16N)}
\right),\label{eq:phasedep1}
\end{equation}
where $\sin(\phi/N)$ has been approximated by $\phi/N$.
Numerical diagonalisation clearly confirms the dependence on
$\tilde D_N^{(3)}$, although a numerical correction factor
$b_\phi$ of the order of $0.7$--$0.75$ for all $N$ has to be added.
Additionally, but expectedly, the phase dependence is 
slowly dampened for larger values of $\Vert \vec x\Vert$.
Yhe magnitude of these amplitudes
rapidly decreases which  makes the imaginary components
even smaller.  Thus, the   leading component of the phase 
simplifies to
\begin{equation}
\exp(i b_\phi \sin(\phi/N)\tilde\omega^2 D_N^{(3)} /12),
\end{equation}
where $b_\phi\approx 0.7$. 
Finally, we turn in the direction
of inhomogeneous array.

\section{Inhomogeneous arrays and renormalisability
\label{sec:inhomog}}

Our main aim is to obtain a wave function similar to $\psi_1$
in the inhomogeneous case at $\phi=0$ and, subsequently, improve
this wave function. Effects due to non-zero $\phi$ are treatable
in principle, but the expression become rather messy and  accuracy
is not that good. It suffices to say that the behaviour of the
eigenenergy corresponds to the effective coupling 
strength $\varepsilon_{\mathrm{J}}\cos(\phi/N)$.

In the inhomogeneous case the charging energy reads
\begin{equation}
E_{\mathrm{C}}\left[\sum_{j=1}^N\frac{v_j^2}{c_j}-\frac1N
\left(\sum_{j=1}^N\frac{v_j}{c_j}\right)^2\right].
\end{equation}
The biasing to zero voltage implies that $\sum_{j=1}^N(v_j/c_j)=0$,
although the above expression is invariant under transformation
$v_j\rightarrow v_j+y$. As each coupling is multiplied by $c_j$,
the second order approximation for the Hamiltonian becomes
\begin{equation}
-\frac12\sum_{j=1}^N\frac{c_j}{\beta^2}\frac{\partial^2}{\partial
(\beta\hat q_j)^2}+\frac{\omega^2\hat q_j^2}{2c_j},
\end{equation}
where $\beta=\sqrt{N/(N-1)}$.
For sufficiently small values of $\omega$ and reasonably homogeneous
arrays the condition $\sum_{j=1}^N(v_j/c_j)=0$ does not vary
much between neighbouring points. In other words, the error between
different lines of the eigenvalue equation is insignificant. 
Under those circumstances we renormalise the coordinates
according to
\begin{equation}
v_j\rightarrow\tilde v_j=v_j/\sqrt{c_j},
\end{equation}
which yields a Hamiltonian identical to the homogeneous case.
In a similar manner, we write the the lowest order
wave function as
\begin{equation}
\psi_1^{(\mathrm{inh})}(\vec x)\propto \exp\left(-\frac{(\omega/2)
\sum_{j=1}^{N} (v_j^2/c_j)}{1-(N-1)\omega/8}\right).\label{eq:inhwfn1}
\end{equation}
where the summation gives simply the charging energy corresponding
to $\vec x$. This is the best Gaussian wave function in the
renormalised coordinates
$\tilde v_j$ and the rate of convergence of the error 
the expected $1/\sqrt{\varepsilon_{\mathrm{J}}}$. 

The Hamiltonian of an inhomogeneous Cooper pair pump is thus 
renormalisable and the leading terms in the eigenenergy are
\begin{equation}
E_{0,\mathrm{inh}}^{(N)}\approx 
-\varepsilon_{\mathrm{J}}\sum_{j=1}^Nc_j+(N-1)\sqrt{\frac
{\varepsilon_{\mathrm{J}}}{2}}+\mathcal{O}(1).\label{eq:inhenergy}
\end{equation}
The constant term can also be evaluated if we assume a
cancellation of $3/4$ in this term which is correct
for homogeneous arrays. We simplify
the expression
\begin{equation}
-\frac{1}{16}\sum_{j=1}^N\frac{1}{c_j}\left[1-1/(Nc_j)\right]^2
\end{equation}
by denoting $b_j:=(1/c_j-1)$ and collecting the terms.
Not so unexpectedly, and as  in Ref.~\onlinecite{aun00}, 
the deviation from the homogeneous value is dominantly
proportional to the square of the inhomogeneity
index $X_{\mathrm{inh}}$. The result,
\begin{equation}
\frac{-(N-1)^2-(2N-3)X_{\mathrm{inh}}^2+\sum_{j=1}^N(b_j^3/N)}{16N},
\end{equation}
has been confirmed up to $N=6$ if only a single capacitance
deviates from the others. In case $N=3$ this expression has 
been tested more rigorously and further corrections 
do vanish as $1/\sqrt{\varepsilon_{\mathrm{J}}}$.

In order to improve the results, more elaborate transformations are 
required. The most viable transformation is based on diagonalising
the charging energy and transforming the representation space 
($\vec x$-space) in such a manner that the charging energy is
proportional to the square of the new norm. New representatives
$\hat q'_j$ are obtained and the differential operators in the
second and fourth order can be obtained. For some special cases, 
the  second order  differential operator is of the Laplace type, 
i.e. the conjugate coordinate is given by
\begin{equation}
\tilde D'_2(\vec x') =\sum_{j=1}^Nc_j(\hat q'_j\cdot\vec x')^2=
\Vert \vec x'\Vert^2.\label{eq:laplace1}
\end{equation}
In those cases the fourth order coordinate
\begin{equation}
\tilde D'_4(\vec x') =\sum_{j=1}^Nc_j(\hat q'_j\cdot\vec x')^4
\end{equation}
yields a trial wave function which can be compared against the 
numerically obtained wave function. In most cases, the
Laplacian operator is slightly distorted, but for small
inhomogeneities this can be neglected as the first
approximation.  In both cases the results  are not as good as in the
homogeneous case, but  the improvement with respect to 
Eq.~(\ref{eq:inhwfn1}) is significant.
Due to dimensional limitations the comparisons between
wave functions have been performed when $N=3$.

As shown by the cancellation in the eigenenergy, 
no isotropic value of $\alpha$ such as $3(N-1)/16N$ in 
Eq.~(\ref{eq:genwfn}) is can be used. Rescaling of the
coordinates changes the optimal value of $\alpha$ in 
different directions, and some further improvement may be obtained 
by using a non-isotropic $\alpha(\vec x)$ in the calculations.
Minor improvements can be obtained by 
fiddling with the coefficients of the coordinates, too.
We conclude this section by stating that significant
improvement of the wave function has been obtained, but
so far no analytical expressions have been able to
reach asymptotical convergence better than  
$1/\sqrt{\varepsilon_{\mathrm{J}}}$.

\section{Conclusions
\label{sec:conclu}}

We have developed a method for obtaining an (approximate)
analytical solution for Laplace type eigenvalue equations
with a harmonic potential and discreteness induced 
higher order corrections. In the one-dimensional case
corresponding to the Mathieu equation the results were
convincing and thus we applied the proposed method on
the tunnelling-charging Hamiltonian of an
ideally biased Cooper pair pump. 

We have obtained reliable analytical expressions for the 
ground state wave function and energy for homogeneous
arrays of arbitrary length. Furthermore, effects due to
nonvanishing phase difference were relatively well described
and the Hamiltonian of an inhomogeneous pump was shown to
be renormalisable. Again, reliable eigenenergies and
reasonable eigenfunctions were obtained. Further 
improvements in the inhomogeneous case have been
proposed and partially carried out, too.

\begin{acknowledgments}
This work has been supported by the Academy of Finland
under the Finnish Centre of Excellence Programme 2000-2005
(Project No. 44875, Nuclear and Condensed Matter Programme at JYFL).
The author thanks Mr. J.~J. Toppari for discussions and
help with the figures.
\end{acknowledgments}

\bibliography{LargeejcV2}

\begin{thebibliography}{31}
\expandafter\ifx\csname natexlab\endcsname\relax\def\natexlab#1{#1}\fi
\expandafter\ifx\csname bibnamefont\endcsname\relax
  \def\bibnamefont#1{#1}\fi
\expandafter\ifx\csname bibfnamefont\endcsname\relax
  \def\bibfnamefont#1{#1}\fi
\expandafter\ifx\csname citenamefont\endcsname\relax
  \def\citenamefont#1{#1}\fi
\expandafter\ifx\csname url\endcsname\relax
  \def\url#1{\texttt{#1}}\fi
\expandafter\ifx\csname urlprefix\endcsname\relax\def\urlprefix{URL }\fi
\providecommand{\bibinfo}[2]{#2}
\providecommand{\eprint}[2][]{\url{#2}}

\bibitem[{\citenamefont{Averin}(1998)}]{ave98}
\bibinfo{author}{\bibfnamefont{D.~V.} \bibnamefont{Averin}},
  \bibinfo{journal}{Solid State Commun.} \textbf{\bibinfo{volume}{105}},
  \bibinfo{pages}{659} (\bibinfo{year}{1998}).

\bibitem[{\citenamefont{Pekola et~al.}(1999)\citenamefont{Pekola, Toppari,
  Savolainen, and Averin}}]{pek99}
\bibinfo{author}{\bibfnamefont{J.~P.} \bibnamefont{Pekola}},
  \bibinfo{author}{\bibfnamefont{J.~J.} \bibnamefont{Toppari}},
  \bibinfo{author}{\bibfnamefont{M.~T.} \bibnamefont{Savolainen}},
  \bibnamefont{and} \bibinfo{author}{\bibfnamefont{D.~V.}
  \bibnamefont{Averin}}, \bibinfo{journal}{Phys.~Rev.~B}
  \textbf{\bibinfo{volume}{60}}, \bibinfo{pages}{R9931} (\bibinfo{year}{1999}).

\bibitem[{\citenamefont{Makhlin et~al.}(1999)\citenamefont{Makhlin,
  Sch$\ddot{\mathrm{o}}$n, and Shnirman}}]{mak99}
\bibinfo{author}{\bibfnamefont{Y.}~\bibnamefont{Makhlin}},
  \bibinfo{author}{\bibfnamefont{G.}~\bibnamefont{Sch$\ddot{\mathrm{o}}$n}},
  \bibnamefont{and} \bibinfo{author}{\bibfnamefont{A.}~\bibnamefont{Shnirman}},
  \bibinfo{journal}{Nature} \textbf{\bibinfo{volume}{386}},
  \bibinfo{pages}{305} (\bibinfo{year}{1999}).

\bibitem[{\citenamefont{Averin}(2000)}]{ave00}
\bibinfo{author}{\bibfnamefont{D.~V.} \bibnamefont{Averin}}, in
  \emph{\bibinfo{booktitle}{Exploring the quantum-classical frontier}}, edited
  by \bibinfo{editor}{\bibfnamefont{J.~R.} \bibnamefont{Friedman}}
  \bibnamefont{and} \bibinfo{editor}{\bibfnamefont{S.}~\bibnamefont{Han}}
  (\bibinfo{publisher}{Nova science publishers}, \bibinfo{address}{Commack,
  NY}, \bibinfo{year}{2000}), \bibinfo{note}{cond-mat/0004364}.

\bibitem[{\citenamefont{Falci et~al.}(2000)\citenamefont{Falci, Fazio, Palma,
  Siewert, and Vedral}}]{fal00}
\bibinfo{author}{\bibfnamefont{G.}~\bibnamefont{Falci}},
  \bibinfo{author}{\bibfnamefont{R.}~\bibnamefont{Fazio}},
  \bibinfo{author}{\bibfnamefont{G.~H.} \bibnamefont{Palma}},
  \bibinfo{author}{\bibfnamefont{J.}~\bibnamefont{Siewert}}, \bibnamefont{and}
  \bibinfo{author}{\bibfnamefont{V.}~\bibnamefont{Vedral}},
  \bibinfo{journal}{Nature} \textbf{\bibinfo{volume}{407}},
  \bibinfo{pages}{355} (\bibinfo{year}{2000}).

\bibitem[{\citenamefont{Pekola and Toppari}(2001)}]{pek01}
\bibinfo{author}{\bibfnamefont{J.~P.} \bibnamefont{Pekola}} \bibnamefont{and}
  \bibinfo{author}{\bibfnamefont{J.~J.} \bibnamefont{Toppari}},
  \bibinfo{journal}{Phys.~Rev.~B} \textbf{\bibinfo{volume}{64}},
  \bibinfo{pages}{172509} (\bibinfo{year}{2001}).

\bibitem[{\citenamefont{Mooij et~al.}(1999)\citenamefont{Mooij, Orlando,
  Levitov, Tian, van~der Val, and Lloyd}}]{moo99}
\bibinfo{author}{\bibfnamefont{J.~E.} \bibnamefont{Mooij}},
  \bibinfo{author}{\bibfnamefont{T.~P.} \bibnamefont{Orlando}},
  \bibinfo{author}{\bibfnamefont{L.}~\bibnamefont{Levitov}},
  \bibinfo{author}{\bibfnamefont{L.}~\bibnamefont{Tian}},
  \bibinfo{author}{\bibfnamefont{C.~H.} \bibnamefont{van~der Val}},
  \bibnamefont{and} \bibinfo{author}{\bibfnamefont{S.}~\bibnamefont{Lloyd}},
  \bibinfo{journal}{Science} \textbf{\bibinfo{volume}{285}},
  \bibinfo{pages}{1036} (\bibinfo{year}{1999}).

\bibitem[{\citenamefont{Nakamura et~al.}(1999)\citenamefont{Nakamura, Pashkin,
  and Tsai}}]{nak99}
\bibinfo{author}{\bibfnamefont{Y.}~\bibnamefont{Nakamura}},
  \bibinfo{author}{\bibfnamefont{Y.~A.} \bibnamefont{Pashkin}},
  \bibnamefont{and} \bibinfo{author}{\bibfnamefont{J.~S.} \bibnamefont{Tsai}},
  \bibinfo{journal}{Nature} \textbf{\bibinfo{volume}{398}},
  \bibinfo{pages}{786} (\bibinfo{year}{1999}).

\bibitem[{\citenamefont{Orlando et~al.}(1999)\citenamefont{Orlando, Mooij,
  Tian, van~der Val L.~Levitov, , and Lloyd}}]{orl99}
\bibinfo{author}{\bibfnamefont{T.~P.} \bibnamefont{Orlando}},
  \bibinfo{author}{\bibfnamefont{J.~E.} \bibnamefont{Mooij}},
  \bibinfo{author}{\bibfnamefont{L.}~\bibnamefont{Tian}},
  \bibinfo{author}{\bibfnamefont{C.~H.} \bibnamefont{van~der Val L.~Levitov}},
  , \bibnamefont{and} \bibinfo{author}{\bibfnamefont{S.}~\bibnamefont{Lloyd}},
  \bibinfo{journal}{Phys.~Rev.~B.} \textbf{\bibinfo{volume}{60}},
  \bibinfo{pages}{15398} (\bibinfo{year}{1999}).

\bibitem[{\citenamefont{Choi et~al.}(2001)\citenamefont{Choi, Fazio, Siewert,
  and Bruder}}]{cho01}
\bibinfo{author}{\bibfnamefont{M.~S.} \bibnamefont{Choi}},
  \bibinfo{author}{\bibfnamefont{R.}~\bibnamefont{Fazio}},
  \bibinfo{author}{\bibfnamefont{J.}~\bibnamefont{Siewert}}, \bibnamefont{and}
  \bibinfo{author}{\bibfnamefont{C.}~\bibnamefont{Bruder}},
  \bibinfo{journal}{Europhys. Lett.} \textbf{\bibinfo{volume}{53}},
  \bibinfo{pages}{251} (\bibinfo{year}{2001}).

\bibitem[{\citenamefont{Bibow et~al.}(2002)\citenamefont{Bibow, Lafarge, and
  Levy}}]{bib02}
\bibinfo{author}{\bibfnamefont{E.}~\bibnamefont{Bibow}},
  \bibinfo{author}{\bibfnamefont{P.}~\bibnamefont{Lafarge}}, \bibnamefont{and}
  \bibinfo{author}{\bibfnamefont{L.~P.} \bibnamefont{Levy}},
  \bibinfo{journal}{Phys.~Rev.~Lett} \textbf{\bibinfo{volume}{88}},
  \bibinfo{pages}{017003} (\bibinfo{year}{2002}).

\bibitem[{\citenamefont{Makhlin et~al.}(2001)\citenamefont{Makhlin,
  Sch$\ddot{\mathrm{o}}$n, and Shnirman}}]{mak01}
\bibinfo{author}{\bibfnamefont{Y.}~\bibnamefont{Makhlin}},
  \bibinfo{author}{\bibfnamefont{G.}~\bibnamefont{Sch$\ddot{\mathrm{o}}$n}},
  \bibnamefont{and} \bibinfo{author}{\bibfnamefont{A.}~\bibnamefont{Shnirman}},
  \bibinfo{journal}{Rev.~Mod.~Phys.} \textbf{\bibinfo{volume}{73}},
  \bibinfo{pages}{357} (\bibinfo{year}{2001}).

\bibitem[{\citenamefont{Hassel and Sepp$\ddot{\mathrm{a}}$}(1999)}]{has99}
\bibinfo{author}{\bibfnamefont{J.}~\bibnamefont{Hassel}} \bibnamefont{and}
  \bibinfo{author}{\bibfnamefont{H.}~\bibnamefont{Sepp$\ddot{\mathrm{a}}$}}
  (\bibinfo{year}{1999}), \bibinfo{note}{in Proc. of 22$^{\mathrm{nd}}$ Int.
  Conf. on Low Temp. Phys.}

\bibitem[{\citenamefont{Aunola et~al.}(2000)\citenamefont{Aunola, Toppari, and
  Pekola}}]{aun00}
\bibinfo{author}{\bibfnamefont{M.}~\bibnamefont{Aunola}},
  \bibinfo{author}{\bibfnamefont{J.~J.} \bibnamefont{Toppari}},
  \bibnamefont{and} \bibinfo{author}{\bibfnamefont{J.~P.}
  \bibnamefont{Pekola}}, \bibinfo{journal}{Phys.~Rev.~B}
  \textbf{\bibinfo{volume}{62}}, \bibinfo{pages}{1296} (\bibinfo{year}{2000}).

\bibitem[{\citenamefont{Aunola}(2001)}]{aun01}
\bibinfo{author}{\bibfnamefont{M.}~\bibnamefont{Aunola}},
  \bibinfo{journal}{Phys.~Rev.~B} \textbf{\bibinfo{volume}{63}},
  \bibinfo{pages}{132508} (\bibinfo{year}{2001}).

\bibitem[{\citenamefont{Abramowitz and Stegun}(1972)}]{abr72}
\bibinfo{author}{\bibfnamefont{M.}~\bibnamefont{Abramowitz}} \bibnamefont{and}
  \bibinfo{author}{\bibfnamefont{I.~A.} \bibnamefont{Stegun}},
  \emph{\bibinfo{title}{Handbook of Mathematical Functions}}
  (\bibinfo{publisher}{Dover}, \bibinfo{address}{New York},
  \bibinfo{year}{1972}).

\bibitem[{\citenamefont{Eiles and Martinis}(1994)}]{eil94}
\bibinfo{author}{\bibfnamefont{T.~M.} \bibnamefont{Eiles}} \bibnamefont{and}
  \bibinfo{author}{\bibfnamefont{J.~M.} \bibnamefont{Martinis}},
  \bibinfo{journal}{Phys.~Rev. B} \textbf{\bibinfo{volume}{64}},
  \bibinfo{pages}{R627} (\bibinfo{year}{1994}).

\bibitem[{\citenamefont{Tinkham}(1996)}]{tink96}
\bibinfo{author}{\bibfnamefont{M.}~\bibnamefont{Tinkham}},
  \emph{\bibinfo{title}{Introduction to superconductivity, 2$^{\mathrm{nd}}$
  ed.}} (\bibinfo{publisher}{McGraw-Hill}, \bibinfo{address}{New York},
  \bibinfo{year}{1996}), pp. \bibinfo{pages}{274--277}.

\bibitem[{\citenamefont{Skeel and Hardy}(2001)}]{ske01}
\bibinfo{author}{\bibfnamefont{R.~D.} \bibnamefont{Skeel}} \bibnamefont{and}
  \bibinfo{author}{\bibfnamefont{D.~J.} \bibnamefont{Hardy}},
  \bibinfo{journal}{SIAM J. Sc. Comp.} \textbf{\bibinfo{volume}{23}},
  \bibinfo{pages}{1172} (\bibinfo{year}{2001}).

\bibitem[{\citenamefont{Reich}(1999)}]{rei99}
\bibinfo{author}{\bibfnamefont{S.}~\bibnamefont{Reich}}, \bibinfo{journal}{SIAM
  J.~Numer.~Anal.} \textbf{\bibinfo{volume}{36}}, \bibinfo{pages}{1549}
  (\bibinfo{year}{1999}).

\bibitem[{\citenamefont{Gans and Shalloway}(2000)}]{gan00}
\bibinfo{author}{\bibfnamefont{J.}~\bibnamefont{Gans}} \bibnamefont{and}
  \bibinfo{author}{\bibfnamefont{D.}~\bibnamefont{Shalloway}},
  \bibinfo{journal}{Phys.~Rev.~E} \textbf{\bibinfo{volume}{61}},
  \bibinfo{pages}{4587} (\bibinfo{year}{2000}).

\bibitem[{\citenamefont{Hairer and Lubich}(2000)}]{hai00}
\bibinfo{author}{\bibfnamefont{E.}~\bibnamefont{Hairer}} \bibnamefont{and}
  \bibinfo{author}{\bibfnamefont{C.}~\bibnamefont{Lubich}}, in
  \emph{\bibinfo{booktitle}{Dynamics of algorithms}}, edited by
  \bibinfo{editor}{\bibfnamefont{R.}~\bibnamefont{de~la Llave}},
  \bibinfo{editor}{\bibfnamefont{L.}~\bibnamefont{Petzold}}, \bibnamefont{and}
  \bibinfo{editor}{\bibfnamefont{J.}~\bibnamefont{Lorenz}}
  (\bibinfo{publisher}{Springer}, \bibinfo{address}{Heidelberg},
  \bibinfo{year}{2000}), p.~\bibinfo{pages}{91}, \bibinfo{note}{volume 118 in
  series Volumes in mathematics and its applications}.

\bibitem[{\citenamefont{Ingold and Nazarov}(1992)}]{ing92}
\bibinfo{author}{\bibfnamefont{G.-L.} \bibnamefont{Ingold}} \bibnamefont{and}
  \bibinfo{author}{\bibfnamefont{Y.~V.} \bibnamefont{Nazarov}}, in
  \emph{\bibinfo{booktitle}{Single charge tunnelling, Coulomb blockade
  phenomena in nanostructures}}, edited by
  \bibinfo{editor}{\bibfnamefont{H.}~\bibnamefont{Grabert}} \bibnamefont{and}
  \bibinfo{editor}{\bibfnamefont{M.~L.} \bibnamefont{Devoret}}
  (\bibinfo{publisher}{Plenum}, \bibinfo{address}{New York},
  \bibinfo{year}{1992}), chap.~\bibinfo{chapter}{2}.

\bibitem[{\citenamefont{Rudin}(1973)}]{rud73}
\bibinfo{author}{\bibfnamefont{W.}~\bibnamefont{Rudin}},
  \emph{\bibinfo{title}{Functional analysis}}
  (\bibinfo{publisher}{McGraw-Hill}, \bibinfo{address}{New York},
  \bibinfo{year}{1973}).

\bibitem[{\citenamefont{Berry}(1984)}]{ber84}
\bibinfo{author}{\bibfnamefont{M.~V.} \bibnamefont{Berry}},
  \bibinfo{journal}{Proc. R. Soc. London, Ser. A}
  \textbf{\bibinfo{volume}{392}}, \bibinfo{pages}{45} (\bibinfo{year}{1984}).

\bibitem[{\citenamefont{Aunola}(2002)}]{aun02}
\bibinfo{author}{\bibfnamefont{M.}~\bibnamefont{Aunola}}
  (\bibinfo{year}{2002}), \bibinfo{note}{cond-mat/0203440 (unpublished)}.

\bibitem[{\citenamefont{Mei$\beta$ner and Steinborn}(1999)}]{mei99}
\bibinfo{author}{\bibfnamefont{H.}~\bibnamefont{Mei$\beta$ner}}
  \bibnamefont{and} \bibinfo{author}{\bibfnamefont{E.~O.}
  \bibnamefont{Steinborn}}, \bibinfo{journal}{Phys.~Rev.~A}
  \textbf{\bibinfo{volume}{56}}, \bibinfo{pages}{1189} (\bibinfo{year}{1999}).

\bibitem[{\citenamefont{Meurice}(2002)}]{meu02}
\bibinfo{author}{\bibfnamefont{Y.}~\bibnamefont{Meurice}}
  (\bibinfo{year}{2002}), \bibinfo{note}{quant-ph/0202047 (unpublished)}.

\bibitem[{\citenamefont{Aunola}(2003{\natexlab{a}})}]{aun03}
\bibinfo{author}{\bibfnamefont{M.}~\bibnamefont{Aunola}},
  \bibinfo{journal}{J.~Math.~Phys.} \textbf{\bibinfo{volume}{44}},
  \bibinfo{pages}{1913} (\bibinfo{year}{2003}{\natexlab{a}}).

\bibitem[{\citenamefont{Aunola}(2003{\natexlab{b}})}]{aun03b}
\bibinfo{author}{\bibfnamefont{M.}~\bibnamefont{Aunola}}
  (\bibinfo{year}{2003}{\natexlab{b}}), \bibinfo{note}{math-ph/0304041
  (unpublished)}.

\bibitem[{tek()}]{teksti}
\bibinfo{note}{It seems reasonable to assume that the ansatz works for larger
  values of $N$, too. The calculations can be repeated using attached
  \textsc{MATHEMATICA} notebook \texttt{GSsolutions.nb} which is available with
  the \TeX\ source.}

\end{thebibliography}

\end{document}